\documentclass[prl,aps,twocolumn, showpacs,amsmath,amssymb]{revtex4}
\usepackage{epsf}
\begin{document}
\title{Hanle effect driven by weak-localization}
\author{I.~S.~Lyubinskiy,  V.~Yu.~Kachorovskii}
\affiliation{A.F.~Ioffe Physical-Technical Institute, 26
Polytechnicheskaya str., Saint Petersburg, 194021, Russia}
\date{\today}
\begin{abstract}
{ The influence of weak localization on Hanle effect
 in a two-dimensional  system with spin-split spectrum is considered.
We show that  weak localization drastically changes the dependence
of stationary spin polarization $\mathbf S$ on external magnetic
field $B.$ In particular, the non-analytic dependence of $\mathbf
S$ on $\mathbf B$ is predicted for III-V-based quantum wells grown
in [110] direction and for [100]-grown quantum wells having  equal
strengths of Dresselhaus and  Bychkov-Rashba spin-orbit coupling.
It is shown that in weakly localized regime the  components of
$\mathbf S$ are
  discontinuous  at $B=0.$
  At  low $B,$ the magnetic
  field-induced rotation of the stationary polarization is
  determined by quantum interference effects.
 This implies that
   the Hanle effect in such  systems
 is totally driven by weak localization.}
\end{abstract}
 \maketitle
\vspace{-0.3cm}
 In recent years, the spin-related phenomena in semiconductor nanostructures have been
 subject to intense study
 both in terms of
fundamental physics and in view of  applications in the field of
spintronics \cite{avsh}. The ultimate goal of spintronics is to
develop novel electronic devices that exploit the spin degree of
freedom. The effective manipulation of the spin in such devices
requires that the characteristic spin lifetime be  long compared
to the device operation time. This is a challenging problem,
especially for III-V-based semiconductor nanostructures where spin
polarization relaxes rapidly due to Dyakonov-Perel  spin
relaxation mechanism \cite{perel}. This mechanism is based on the
classical picture of the  angular spin diffusion in random
magnetic field induced by spin-orbit coupling. In two-dimensional
(2D) systems, the corresponding spin-relaxation time $\tau_S$ is
inversely proportional to the momentum relaxation time $\tau$
\cite{dyak}.
 As a consequence,  in high-mobility structures which are most promising for
device applications, $\tau_S$ is especially short. However,
 in some special cases, the relaxation of one of the spin
components can be rather slow even in a system with high mobility.
 In particular,  a number of recent researches
\cite{d1,d2,d3,d4,d5,d6} are devoted to GaAs symmetric quantum
wells (QW) grown in $[110]$ direction. In such wells, the random
field is  perpendicular to the QW plane and the normal to the
plane component of the spin does not relax \cite{dyak} or, more
precisely, relaxes very slowly. Also, the random magnetic field
might be parallel to a fixed axis in an asymmetric $[100]$-grown
QW due to the interplay between structural and bulk spin-orbit
coupling \cite{pikus,golub,kim,loss1,loss2}. Since in both cases
one component of the spin  relaxes slowly, these structures are
especially attractive for spintronics applications.

In this Letter, we discuss the dependence of the spin polarization
in such structures on external magnetic field $\mathbf B.$ We
assume that the spin is injected into the system  with a constant
rate, for example, by optical excitation \cite{optical}. The
stationary spin polarization $\mathbf S $ is proportional to the
product of injection intensity   and the spin relaxation time.
 The Hanle effect is that the external field modifies the
stationary polarization. In particular, $\mathbf S(\mathbf B)$
deviates from $\mathbf S(0)$ by an angle $\theta$ which depends on
the relation between spin precession frequency
$\boldsymbol{\Omega}=g \mu_B \mathbf B/\hbar,$ and the spin
relaxation rate (here $\mu_B=e\hbar/2m_0c$ is the Bohr magneton,
$m_0$ is the free electron mass  and $g$ is Land\'e $g$ factor).
We show that at low temperatures $\theta$ is very sensitive to
weak localization (WL) effects. Usually such effects are discussed
in context of quantum corrections to the conductivity
\cite{wl,lee}. Though the WL correction is small compared to the
classical conductivity it has attracted much attention due to its
fundamental nature and anomalous behavior with external parameters
such as magnetic field. Remarkably,   the influence of WL  on
Hanle effect can not be considered as a small correction to the
classical result. We demonstrate that  at low temperatures WL
gives rise to a discontinuity in the dependence of $\theta$ on
$\mathbf B.$
 As a
result, at low  $B$  the Hanle effect is totally driven by WL.

\begin{figure}
 \centerline{ {\epsfxsize=6.84cm{\epsfbox{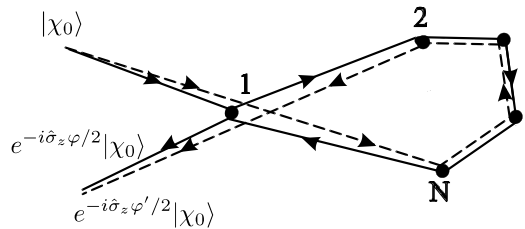}} }}
\vspace{-6mm} \vspace{11mm} \caption{ A coherent scattering
involving $N$ impurities. At point $1$ the electron wave splits
into two parts propagating around the closed loop clockwise and
counterclockwise. The initial electron  spinor $|\chi_0\rangle$ is
transformed to $|\chi\rangle=e^{-i\hat \sigma_z \varphi
/2}|\chi_0\rangle$ and $|\chi'\rangle=e^{-i\hat \sigma_z \varphi'
/2}|\chi\rangle$  for clockwise and counterclockwise paths,
respectively. } \label{fig1}
\end{figure}

Physically, the non-analytic dependence of $\theta$ on $\mathbf B$
is related to memory effects specific for WL. It is known that the
WL is caused by interference of electron waves  propagating along
a closed loop in the opposite directions \cite{lee}. Such
interference process can be considered as a coherent scattering
(additional  to the Born scattering)    involving a large number
of impurities \cite{self} (see Fig.~\ref{fig1}). The probability
of  the coherent scattering is proportional to the probability of
the diffusive return $1/Dt,$ where  $t$ is the time of the
electron passage along the loop and $D$ is the diffusion
coefficient. In the absence of magnetic field, such scattering
does not change the direction of the spin \cite{my}. Thus,
electrons keep memory about initial spin polarization during the
time much larger than $\tau_S$ and the long-living tail $1/t$ in
the spin polarization appears \cite{my,m2}. When the external
magnetic field is applied, the electron spin rotates with a
frequency $\boldsymbol{\omega}_{\mathbf p} + \boldsymbol{\Omega},$
where $\boldsymbol{\omega}_{\mathbf p}$ is the momentum-dependent
frequency  of the spin precession in the spin-orbit-induced
magnetic field.  In the special case under discussion,
$\boldsymbol{\omega}_{\mathbf p}$ is parallel to a fixed axis (say
$z$-axis) for any $\mathbf p$. Let us consider the simplest case,
when $\boldsymbol{\Omega} $ also lies along  $z$ axis. In this
case, spin rotation matrices, describing rotation of electron spin
on the different segments of the closed loop commutes with each
other and the spin rotation angles for clockwise and
counterclockwise propagating waves are simply given by $\varphi =
\Omega t + \int_0^t \omega_{\mathbf p}dt',~~\varphi' = \Omega t +
\int_0^t \omega_{\mathbf p'}dt'.$ The initial electron spinor
$|\chi_0\rangle$ is transformed to $|\chi\rangle=e^{-i\hat
\sigma_z \varphi /2}|\chi_0\rangle$ and $|\chi'\rangle=e^{-i\hat
\sigma_z \varphi' /2}|\chi_0\rangle$ for clockwise and
counterclockwise paths, respectively. Here $\hat \sigma_z$ is the
Pauli matrix. In 2D systems, $\boldsymbol{\omega}_{\mathbf p}$ is
linear in $\mathbf p.$ Since for any closed path $\int_0^t \mathbf
p(t')dt'=0,$ we find $\varphi =\varphi' =\Omega t.$ The
interference contribution to the electron spin-density matrix
after a coherent scattering is given by $|\chi \rangle \langle
\chi'| =e^{-i\hat \sigma_z \Omega t /2}|\chi_0 \rangle \langle
\chi_0| e^{i\hat \sigma_z \Omega t/2}.$ Hence, the electron spin
after the coherent scattering is rotated by an angle $\Omega t$
with respect to the spin before scattering. The deviation of the
spin from the initial direction is proportional to $\sin(\Omega
t).$ The integration over $t$ weighted with the probability of the
coherent scattering  yields $\theta \sim \int dt \sin(\Omega t)/t
\sim \Omega/\vert \Omega \vert. $ In other words, the rotation
frequency $\Omega$ should be multiplied by the effective rotation
time $1/\vert \Omega \vert, $ which is very long for small
$\Omega.$ Rigorous calculations (see below) give an additional
factor $\lambda/l$ in this result, which reflects the quantum
nature of the phenomena: $ \theta\sim \lambda \Omega/l\vert \Omega
\vert   $ (here $\lambda$ is the electron wave length and $l$ is
the mean free path). In contrast to coherent scattering, the
classical rotation of spin is limited by the spin relaxation time
$\tau_S,$ so that the classical contribution is
 $\theta \sim \Omega \tau_S.$ For
$\Omega \tau_S\ll \lambda/l,$ this  contribution can be neglected
and the Hanle effect is totally driven by WL. The discontinuity at
$\Omega=0$ is smeared  by inelastic scattering, which destroys
phase coherence between clockwise and counterclockwise propagating
paths, thus limiting a time of the coherent spin rotation:
$t<\tau_\varphi,$ where $\tau_{\varphi} $ is the phase breaking
time. It worth noting that the above considerations do not work if
$\boldsymbol{\omega}_{\mathbf p}$ changes its direction with a
change   of $\mathbf p.$ The reason is that the rotation matrixes
corresponding to the different segments of the loop do not commute
in this case. As a result, $\varphi$  and $\varphi'$ no longer
equal to each other and depend not only on the total time spent on
the loop but also on the positions of impurities $1,2,\cdots ,N.$
In such a situation, the external magnetic field leads to the spin
decay rather than to the spin rotation \cite{my}.

 Next we develop a rigorous theory of Hanle effect in the  weakly
localized regime. The Hamiltonian of a 2D system with a spin-split
spectrum is given by
\begin{equation}
H=\frac{\mathbf
p^2}{2m}+\frac{\hbar}{2}[\boldsymbol{\omega}_{\mathbf
p}+\boldsymbol{\Omega}]\hat{\boldsymbol{\sigma}} +U(\mathbf r).
\label{hamilt}
\end{equation}
Here $\mathbf p$ is the  in-plane electron momentum, $m$ is the
electron effective mass,
 $\hat{\boldsymbol {\sigma}}$ is a vector consisting of  Pauli matrices,
 and $U(\mathbf r)$  is the impurity potential, which we assume to be short-ranged
 $ \overline{ U(\mathbf r) U(\mathbf r')} = \gamma \delta (\mathbf r -\mathbf r')$ (here averaging
 is taken over impurity positions and the
coefficient  $\gamma$ is related to the transport scattering time
by  $\tau= \hbar^3/m\gamma $). The spin-orbit interaction is
described by the  term $\hbar \boldsymbol{\omega}_{\mathbf
p}\hat{\boldsymbol{\sigma}}/2.$ It can be separated into two parts
($\boldsymbol{\omega}_{\mathbf p}=\boldsymbol{\omega}^1_{\mathbf
p}+\boldsymbol{\omega}^2_{\mathbf p}$)
 related to so-called Dresselhaus \cite{dress} and
Bychkov-Rashba \cite{rashba} contributions. The Bychkov-Rashba
coupling depends on the asymmetry of the QW confining potential.
Its  strength  can be tuned by varying the gate voltage
\cite{nitta}. The Dresselhaus term is present in semiconductors
with no bulk inversion symmetry. In 2D case  the resulting
spectrum splitting is linear in the electron momentum
 \cite{rashba,dyak}.
We assume that the random magnetic field is directed along z-axis
\begin{equation}
\boldsymbol{\omega}_{\mathbf p}=(\mathbf p \boldsymbol{ \alpha})
\hat {\mathbf z}. \label{omegaxi}
\end{equation}
Here $\hat {\mathbf z}$ is the unit vector along the direction of
the random field and $\boldsymbol{ \alpha}$ is a constant in-plain
vector. Equation \eqref{omegaxi} implies that the spectrum
splitting  depends only on one component of momentum $p_{\alpha}.$
This happens in symmetric [110]-grown QWs. In this case, $\hat{
\mathbf z}$ is normal to the QW plane \cite{dyak}. For asymmetric
[100] wells Eq.~\eqref{omegaxi} can also take place if
Bychkov-Rashba and Dresselhaus couplings have equal strengths
\cite{pikus,golub,kim,loss1,loss2}. For such wells vector $\hat{
\mathbf z}$ lies in the QW plane. The temperature is assumed to be
low, so that $\tau_{\varphi}$ be  large compared to $\tau_{S}.$

To describe spin dynamics in the weakly localized regime we use
the kinetic equation \cite{my}. If the spin polarization is
uniform in space, this equation looks as follows
\begin{equation}
\frac{\partial\mathbf s}{\partial t} = \left [ \boldsymbol
{\omega}_{\mathbf p } + \boldsymbol {\Omega} \right ] \times
\mathbf s -\frac{\mathbf s_{\rm a}}{\tau} +\delta\hat J\mathbf s +
\mathbf I, \label{boltz10}
\end{equation}
where $\mathbf s=\mathbf s(\mathbf p)$ is the spin density in the
momentum space, related to the averaged spin by $\mathbf S =
\int\mathbf s d^2 \mathbf p/(2\pi)^2$, $\mathbf s_{\rm a} =\mathbf
s - \langle \mathbf s \rangle $ is the anisotropic part of the
spin density, $\mathbf s_{i}=\langle \mathbf s \rangle $ is its
isotropic part  (here $ \langle \dots \rangle $ stand for
averaging over directions of the electron momentum), $\mathbf I$
is the constant source, which we assume to be monoenergetic, $I
\sim \delta(E-E_0)$, and $\delta \hat J$ is the WL-induced
correction to the Boltzmann collision integral \cite{my,dfi}. For
$E=E_0,$ this correction reads   (see Eq.(24) in Ref. \cite{my})
\begin{equation}
\delta \hat J \mathbf s=- \frac{\lambda l}{\pi
\tau^2}\int_{-\infty}^t dt' \hat W( t-t') \mathbf s_{\rm a}
(\mathbf p, t'). \label{dJ1}
\end{equation}
Here $\lambda=2\pi\hbar/\sqrt{2mE_0},~~l=\tau\sqrt{2E_0/m}$  and
$\hat W(t)$ is the time-nonlocal scattering kernel \cite{coment1}.
Using Eqs. \eqref{boltz10} and \eqref{dJ1} we find a closed
equation
 for the isotropic spin density in the stationary case
\begin{equation}
\hat \Gamma (1-\hat \Lambda) \mathbf s_{\rm i}-\boldsymbol{\Omega}
\times \mathbf s_{\rm i} =\mathbf I.
 \label{boltzstats}
\end{equation}
Here $\hat \Gamma=\hat\tau_S^{-1}$ is the spin relaxation tensor
(tensor of  inverse relaxation times) given by \cite{dyak}
\begin{equation}
\Gamma_{ik}=[\delta_{ik}\langle \omega_{\mathbf p}^2
\rangle-\langle \omega_{\mathbf p i}
 \omega_{\mathbf p k} \rangle]~\tau,
\label{gamma00}
\end{equation}
  and the matrix
\begin{equation}
\hat \Lambda=\frac{\lambda l}{\pi\tau} \int_{\tau }^{\infty}dt~
\hat W( t)
 \label{lambda}
\end{equation}
describes WL correction ($\Lambda_{ik} \ll 1$) \cite{rot}.  As
follows from Eqs.~\eqref{omegaxi} and \eqref{gamma00}, tensor
$\hat \Gamma$ has two nonzero components:
$\Gamma_{xx}=\Gamma_{yy}=\Gamma=mE_0\alpha^2\tau$ \cite{dyak}. We
will assume  that there exists a slow spin relaxation of the z
component of the spin $\Gamma_{zz}=\epsilon \ll \Gamma.$ Such
relaxation  arises due to the cubic in $\mathbf p$ terms,
neglected in the Hamiltonian \eqref{hamilt}, or due to other
mechanisms of the spin relaxation. In the absence of the external
magnetic field the scattering kernel is given by
$W_{ik}=\delta_{ik}/4 \pi D t$ \cite{my}. Therefore,
$\Lambda_{ik}=(\lambda/2\pi^2 l)\delta_{ik}\ln
(\tau_{\varphi}/\tau),$ where phase breaking time $\tau_{\varphi}$
is taken as an upper cut-off of the integral in
Eq.~\eqref{lambda}.

Next we consider the dependence of $\hat \Lambda$ on the magnetic
field for the case when $\mathbf B$ is  parallel to $z$-axis. We
start with discussing of the QW grown in $[100]$ direction. Since
in this case $\hat z$ lies in the QW plane,  the external field
does not affect the orbital motion of the electrons and the Zeeman
term $\hbar \Omega \hat \sigma_z/2 $ commutes with the
Hamiltonian. As a consequence, the solution of Eq.~\eqref{boltz10}
with $I=0$
 and $B\neq 0$ can be obtained from solution at zero
field:
 $\mathbf s_B(t)= \hat T(\Omega t) \mathbf s_{B= 0}(t),$
where $\hat T$ is $3\times 3$  matrix, describing rotation around
z-axis by the angle $\Omega t$.  In order that $\mathbf s_B(t)$
obeys Eq.~\eqref{boltz10}, the scattering kernel has to be as
follows:
\begin{equation}
 \hat W(t)=  \frac{\hat T(\Omega t)}{4\pi D t}.
\label{scatt}
 \end{equation}
 Equations \eqref{boltzstats},\eqref{lambda} and \eqref{scatt}
allows us to  find the isotropic spin polarization. If $\mathbf I$
is perpendicular to the z-axis, we get
\begin{equation}
\mathbf S= (1+\Lambda_c)\mathbf S_0+\left(\Lambda_s+ \Omega/\Gamma
\right) \hat{\mathbf z} \times \mathbf S_0. \label{statt}
\end{equation}
 Here
$\mathbf S_0= \int \mathbf I d^2\mathbf p/(2\pi)^2\Gamma,~\mathbf
S= \int \mathbf s_{\rm i} d^2\mathbf p/(2\pi)^2,$ and
\begin{align}
\Lambda_c + i \Lambda_s &= \frac{\lambda}{2\pi^2
l}\int_{\tau}^{\infty} \frac{dt}{t}~e^{i\Omega t}
\nonumber \\
& \approx \frac{\lambda}{2 \pi^2 l} \left [\ln
\left(\frac{1}{\vert\Omega\vert
 \tau} \right) + \frac {i\pi}{2} \frac {\Omega }{\vert \Omega \vert}
\right]
 \label{Lambda}
\end{align}
For $\Omega/\Gamma \ll 1,$ the angle between $\mathbf S$ and
$\mathbf S_0$ is
\begin{equation}
\theta(\Omega) \approx \frac{\Omega}{\Gamma} +\frac{\lambda}{ 4\pi
l}\frac {\Omega }{\vert \Omega \vert}. \label{hanle}
\end{equation}
In this equation,  $\Omega/\Gamma$  stands for the classical
contribution, while the second term is due to WL effect.  At small
fields the classical contribution can be neglected. We see that WL
contribution to $\theta$ is a non-analytic  function of the
magnetic field. At zero field the function $\theta(\Omega)$ has a
discontinuity
\begin{equation}
\theta(+0)-\theta(-0)=\lambda/2\pi l. \label{discount}
\end{equation}

  In deriving
Eq.~\eqref{hanle}, we neglected inelastic scattering. Such
scattering destroys phase coherence, thus suppressing WL
contribution. In order to take it into account, we introduce a
factor $\exp(- t/\tau_{\varphi})$ in the integrand in
Eq.~\eqref{Lambda}.
 As a result, we find
\begin{equation}
\theta(\Omega) \approx \frac{\Omega}{\Gamma} +\frac{\lambda}{
2\pi^2 l}\arctan (\Omega \tau_{\varphi}). \label{hanletauf}
\end{equation}
This equation shows that for low temperatures, when
$1/\tau_{\varphi} \ll \Gamma \lambda/l,$ the Hanle effect is still
totally driven by WL at small $\Omega.$ With increasing the
temperature, the WL-induced discontinuity is smeared out.

Let us now  consider the system with symmetric QW grown in $[110]$
direction, assuming again that $\boldsymbol{\Omega} \parallel
\hat{\mathbf z}$ and $~\mathbf I \perp \hat {\mathbf z}.$ The
above considerations can be applied with a slight modification of
Eq.~\eqref{Lambda}. Since in this case $\hat{\mathbf z}$ is normal
to the QW plane, the external magnetic field also effects orbital
motion of the electrons leading to the suppression of WL
\cite{lee}. Including into the integral in Eq.~\eqref{Lambda} an
additional factor $\gamma t/\sinh(\gamma t)$ \cite{chakrob}, which
accounts for destruction of WL by external  field (here
$\gamma=2eBD/\hbar c$), we obtain
\begin{equation}
\theta= \frac{\Omega}{\Gamma}+ \frac{\lambda}{4\pi l} \frac{\Omega
}{\vert\Omega\vert} \tanh{\left(\frac{g\lambda  }{8 l
}~\frac{m}{m_0}\right)}. \label{gg}
\end{equation}

Next we find the stationary polarization for arbitrary directions
of $\mathbf B$ and $\mathbf I$.  We choose $x,y$ axes in such a
way that $\Omega_x=0$. The classical solution ($\hat \Lambda=0$)
reads
\begin{equation}
\mathbf S \approx \left(
\begin{array}{ccc}
1 & -\Omega_z/\Gamma & \Omega_y/\epsilon \\
\Omega_z/\Gamma & 1+\Omega_y^2/\epsilon\Gamma & \Omega_z\Omega_y/\epsilon \Gamma\\
-\Omega_y/\epsilon & \Omega_z\Omega_y/\epsilon \Gamma &
\Gamma/\epsilon
\end{array}
\right) \frac{\Gamma ~\mathbf S_0 }{\Gamma+\Omega_y^2/\epsilon }
\label{matrix}
\end{equation}
(here we assumed $\Omega \ll \Gamma$). To find $\hat \Lambda $ we
write \cite{my}
\begin{equation}
W_{ij}(t)  =  \sum_{\beta\gamma\alpha\theta}
 \langle \gamma| \hat \sigma_i |\beta \rangle W_{\beta\gamma\alpha\theta}(0,
t)\langle \alpha|\hat \sigma_j |\theta\rangle/2 ,
\label{transform}
\end{equation}
where  function $W_{\beta \gamma \alpha \theta}(\mathbf r,t)$
obeys \cite{iord,knap,pikus,my}
\begin{align}
&K_{\beta \beta' \theta \theta'} W_{\beta'\gamma
\alpha\theta'}(\mathbf r,t) =\delta(t) \delta(\mathbf
r)\delta_{\alpha \beta} \delta_{\gamma\theta}, \label{K} \nonumber
\\
&\hat K =\frac{\partial }{\partial t}- D\left(\nabla-2i e \mathbf
A/\hbar c + i m \boldsymbol{\xi}
[ \hat \sigma_z^{(1)}+\hat \sigma_z^{(2)}]/2\right)^2 \nonumber \\
 & +  i\boldsymbol{\Omega}[
 \hat{\boldsymbol{\sigma}}^{(1)}-\hat{\boldsymbol{\sigma}}^{(2)}]/{2}.
\end{align}
Here
 $\boldsymbol{\sigma}^{(1)}_{\beta \beta' \theta
\theta'}=
      \langle\beta|\hat{\boldsymbol{\sigma}}| \beta'\rangle\delta_{\theta \theta'},$
     $\boldsymbol{\sigma}^{(2)}_{\beta \beta' \theta \theta'}=
      \delta_{\beta \beta'}\langle \theta|\hat{\boldsymbol{\sigma}}| \theta'\rangle$
      and $\mathbf A$ is the vector potential of external field.
Using Eqs.~\eqref{lambda}, \eqref{transform} and \eqref{K}  one
can find
\begin{equation}
\hat \Lambda=\frac{\lambda}{2\pi^2 l} \int_{\tau }^{\infty}
  dt~\frac{ \gamma' }{\sinh(\gamma' t)}~
 \hat T(\Omega_{z} t). \label{djj}
\end{equation}
 Here $\gamma'=2e\mathbf B \mathbf n D/\hbar c$ and $\mathbf n$ is the unit vector
 normal to the QW plane.
 It is worth noting that the
rotation matrix, entering Eq.~\eqref{djj}, depends only on
z-component of the external field. Accounting for  WL leads to the
following replacement in Eq.~\eqref{matrix}:
\begin{equation}
\Gamma \to \Gamma (1-\Lambda_c ),~~\Omega_z \to\Omega_z + \Gamma
\Lambda_s,~~ \epsilon \to \epsilon (1-\Lambda_0), \label{replace}
\end{equation}
where $\Lambda_c=(\lambda/2\pi^2 l)\ln (t_c/\tau
),~\Lambda_0=(\lambda/2\pi^2 l)\ln (t_0/\tau )$ and
$\Lambda_s=(\lambda/4\pi l)\tanh(\lambda g m \Omega_z/8l
m_0|\boldsymbol{\Omega}\mathbf n|)$ (here
 $t_0^{-1} \sim
\max[|\gamma'|,\tau^{-1}_{\varphi}],~t_c^{-1}\sim
\max[|\gamma'|,|\Omega_z|].$ Eqs.~\eqref{djj} and \eqref{replace}
are valid both for [110] and [100] orientations provided that
$\Omega_z \gg \Delta \Omega_z \sim \max[
\epsilon,\Omega_y^2/\Gamma,1/\tau_{\varphi}]$.
 One can show that the discontinuity  of  $\Lambda_s$ (at the point
$\Omega =0$ ) is smeared over the frequency interval $\Delta
\Omega_z. $

A case  when $\mathbf I$ is parallel to $y$ axis is of particular
interest. For $\Omega_y \gg \epsilon$ and $\Omega_z \ll \Gamma
\Lambda_s$ we find
\begin{equation}
\theta =\arctan \left(  \frac{ \Gamma \Omega_y \Lambda_s}{\epsilon
\Gamma + \Omega_y^2} \right ). \label{equation}
\end{equation}
We see that $\theta \gg \Lambda_s$ due to the factor $\Gamma
\Omega_y/(\Gamma \epsilon +
 \Omega_y^2) \gg 1.$
For  $\Gamma \Omega_y/(\Gamma \epsilon +\Omega_y^2)\gg l/\lambda$
the z-component of the spin will be larger than y-component.
Hence, the quantum effects might lead to rotation of spin by a
large angle.

In the above calculations we assumed that $\mathbf I$ is
homogenous. For slowly varying   $\mathbf I,$ the derived
equations relate  $\mathbf S(\mathbf r)$ with $\mathbf I(\mathbf
r)$  provided that the spatial scale of inhomogeneity $L$ is large
compared to $\sqrt{D/\Gamma}=1/m\alpha$ \cite{D/G}. One can show
that in the opposite case $ l \ll L \ll 1/m\alpha,$ these
equations also valid relating  $\int d\mathbf r \mathbf S(\mathbf
r)$ with $\int d\mathbf r \mathbf I(\mathbf r).$

Finally, we briefly discuss a possible  experimental realization
of the predicted effect. One of the most efficient ways of the
spin injection is the
  optical excitation of interband transitions
with circular polarized light \cite{optical}. The  observation of
the  weakly localized regime requires that $\tau_{\varphi} \gg
\tau_{S}$. In optical experiments, the phase breaking is due to
both inelastic scattering (caused by electron-phonon and
electron-electron interactions) and recombination of electrons
with holes. The recombination of spin-polarized electrons with
holes is suppressed in a n-type highly doped QW  excited by
low-intensity light. In this case, the number of spin polarized
electrons is small compared to their total number  and the  holes
are more probably to recombine with unpolarized electrons.
Therefore, the stationary amount of holes is small and the
recombination of polarized electrons with the holes can be
neglected. The characteristic time of inelastic scattering will be
large at low temperatures if the electrons are excited close to
the Fermi level. In such a situation, the emission of the optical
phonons is forbidden and the  phase breaking  is due to
electron-electron collisions. Since the rate of such collisions
decreases with approaching to the Fermi level, $\tau_{\varphi}$
can be tuned to be much longer than $\tau_S$.

To conclude, the theory of Hanle effect in a 2D system   is
developed for the weakly localized regime. At low external
magnetic fields the Hanle effect is totally driven by quantum
interference effects. In the absence of inelastic scattering, the
components of the spin polarizations are
  discontinuous as functions of the external
field.

\begin{acknowledgments}
We are grateful to  K.~V. Kavokin for useful  discussions.
 This work has been supported by  RFBR, a grant of the
RAS, a grant of the Russian Scientific School 2192.2003.2, and a
grant of the foundation "Dinasty"-ICFPM.
\end{acknowledgments}


\begin{thebibliography}{}
\bibitem{dfi}
 Assuming that  the cyclotron frequency is small compared to
$1/\tau,$ we neglect the term, which describes the momentum
rotation by external field.
\bibitem{coment1} To obtain Eq.~\eqref{dJ1} from Eq.~(24) of Ref.~\cite{my}
we  used the following property $\mathbf s_{\rm a }(-\mathbf p,
t)=-\mathbf s_{\rm a }(\mathbf p, t).$
\bibitem{rot} In deriving of Eq.~\eqref{boltzstats}, we took into account
that $\hat W$ is proportional to the rotation matrix (see
Eqs.~\eqref{scatt}, \eqref{djj}).
\bibitem{D/G} Note that $1/m\alpha$ is a characteristic scale of
non-decaying spin oscillations which might exist in the system
under discussion ( see  Ref.~\cite{loss1}).
\end{thebibliography}
\end{document}